\documentclass[11pt]{article}
\usepackage{epsfig}

\def\cn{\mathop{\textrm{cn}}\nolimits}
\def\sn{\mathop{\textrm{sn}}\nolimits}
\def\dn{\mathop{\textrm{dn}}\nolimits}
\def\sech{\mathop{\textrm{sech}}\nolimits}
\def\sinh{\mathop{\textrm{sinh}}\nolimits}
\def\const{\mathop{\textrm{const}}\nolimits}

\begin{document}
\begin{center}
{\Large \bf Chaotic behaviour of the solutions to a perturbed Korteweg-de Vries equation}\\ \ \\
K.B. Blyuss\\
{\it Department of Mathematics and Statistics\\
University of Surrey, GU2 5XH, Guildford, UK}\\
E-mail: k.blyuss@surrey.ac.uk

\end{center}

\begin{abstract}
Stationary wave solutions of the perturbed Korteweg-de Vries equation
are considered in the presence of external hamiltonian
perturbations. Conditions of their chaotic behaviour are studied with
the help of Melnikov theory. For the homoclinic chaos Poincar\'e
sections are constructed to demonstrate the complicated behaviour, and
Lyapunov exponents are also numerically calculated.

\end{abstract}

\section{Introduction}
\setcounter{equation}{0}
\def\theequation{\arabic{section}.\arabic{equation}}
Recently there has been a significant attention to the chaotic
behaviour of the solutions to partial differential equations. For
example, chaos was found in a perturbed sine-Gordon equation \cite{Z},
nonlinear Schr\"odinger equation \cite{M}, and later for Korteweg-de
Vries-Burgers equation \cite{GT} and its generalization on high-order
nonlinearities and the case of Kadomtsev-Petviashvili equation
\cite{CDPT}. It was shown how chaos can appear in such systems which
were completely integrable without perturbation via the appearance of
subharmonics and homoclinic tangles. For the KdV equation
incorporating dissipation and instability \cite{KT} it was shown that
for the strongly dissipative case the overall evolution of solutions is chaotic with irregular soliton interactions.

In our previous paper \cite{B} we've considered the influence of
time-periodic hamiltonian external perturbations on a KdV system and
showed,  how the stochastic layer can appear on a phase plane near
the unperturbed separatrix. The width of this layer was calculated
with the help of Chirikov criterion for the overlap of resonances. In
this work we continue this study on an example of one-harmonic
perturbation. The fact that this perturbation is taken explicitly
provides the possibility to obtain an exact expression for the
Melnikov function in order to determine the transition to a chaotic
behaviour. To study the system in a near-separatrix region, we
transform the governing ODE into a map for which the width of
stochastic layer can be easily obtained. Finally we plot Poincar\'e
sections of chaotic behaviour and calculate the corresponding Lyapunov
exponents.

We start with the perturbed Korteweg-de Vries equation taken in the form:
\begin{equation}\label{1.1}
u_{t}+cu_{x}+uu_{x}+\beta u_{x x x}=f(u,x-Vt)_{x}.
\end{equation}
\begin{figure}
\hspace{2cm}
\epsfig{width=4cm,file=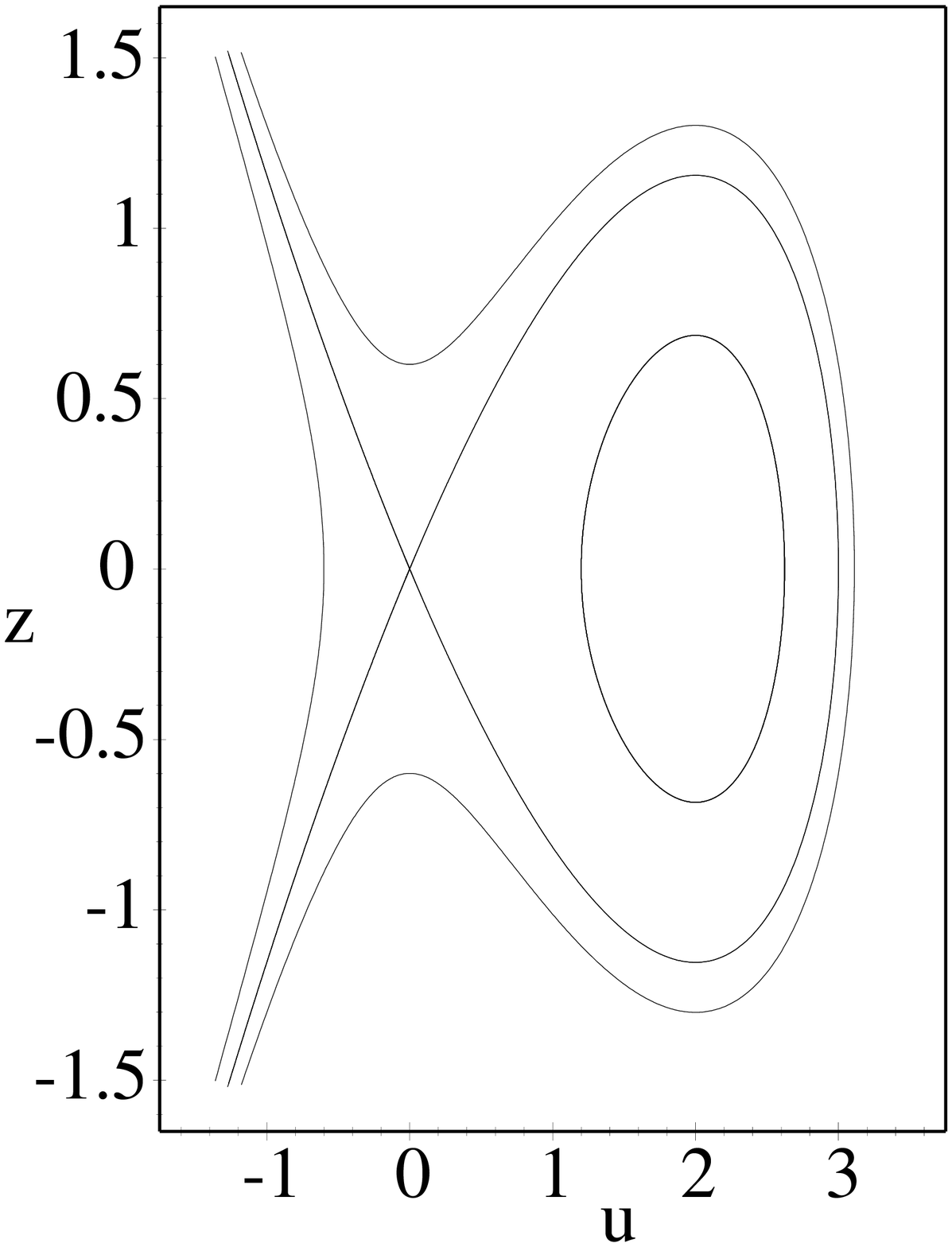}\hspace{2cm}
\epsfig{width=4cm,file=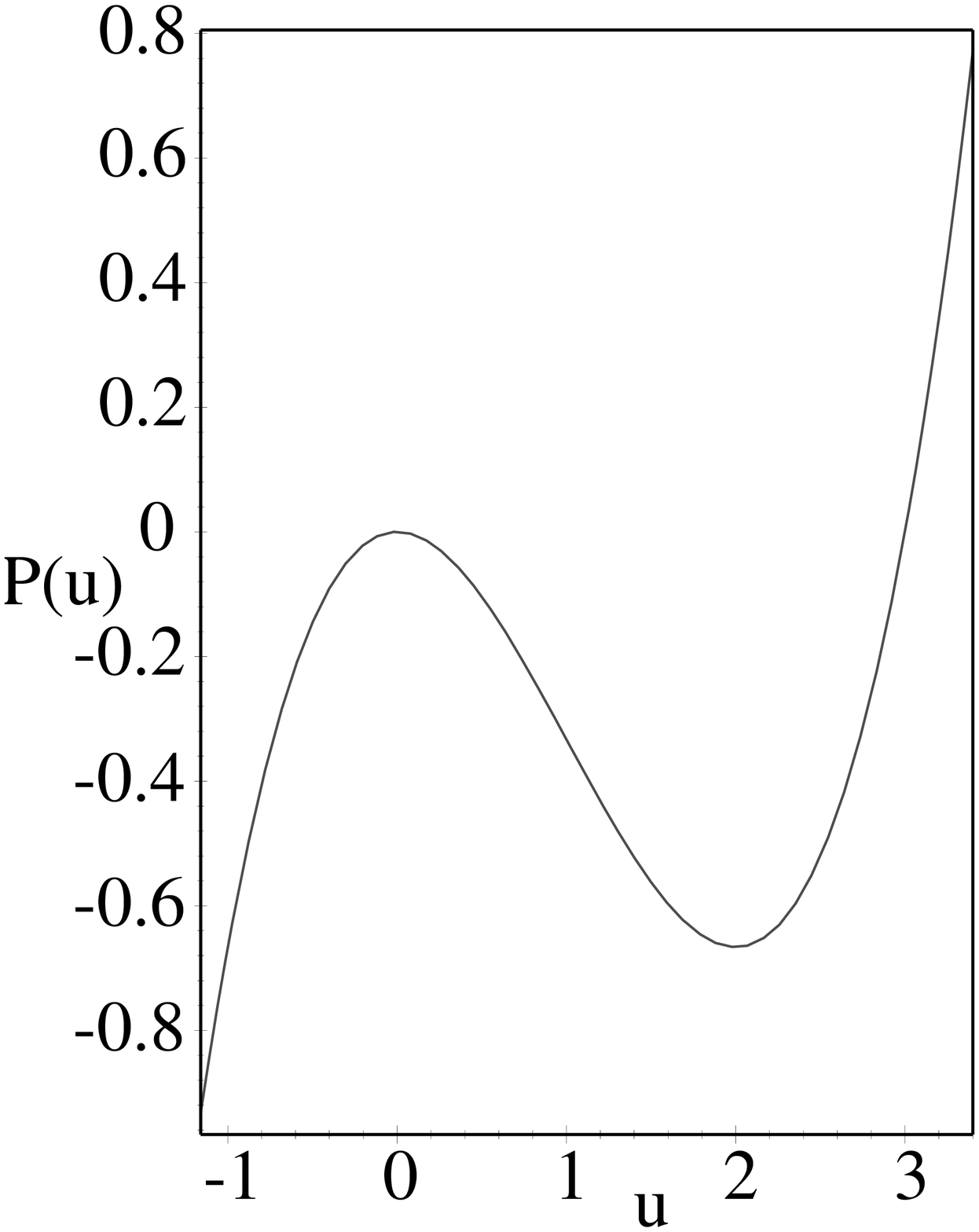}
\caption{ (Left) Phase plane of the system (\ref{2.1}). (Right) Plot of the potential energy (\ref{2.2}). Here $v=1$ and $\beta =1$.}
\end{figure}
Here $f(u,x-Vt)$ is assumed to be periodic on its last argument and will be taken simply as $f=f_{0}\cos\omega(x-Vt)$. Transforming coordinates in a moving frame $(x'\to x-Vt,\mbox{ }t'\to t)$ and considering steady waves $u_{t}=0$, one obtains (the primes are omitted):
\begin{equation}\label{1.3}
\beta u_{x x}=vu-\frac{u^2}{2}+f_{0}\cos\omega x+C,
\end{equation}
where $V=c+v$ (supercritical case) or $V=c-v$ ( subcritical case),
$v>0$, $C$ is the integration constant which can be chosen equal to
zero from the condition that the solution is bounded at
infinity. Without loss of generality we consider here only the supercritical case. Therefore we have:
\begin{equation}\label{1.4}
\beta u_{x x}=vu-\frac{u^2}{2}+f_{0}\cos\omega x
\end{equation}
or with $z\equiv u_{x}$
\begin{equation}\label{1.5}
\left\{
\begin{array}{l}
u_{x}=z\\
z_{x}=\frac{1}{\beta}\left(vu-\frac{u^2}{2}+f_{0}\cos\omega x\right).
\end{array}
\right.
\end{equation}
The outline of the rest of the paper is as follows. In the next
Section the unperturbed case of the Korteweg-de Vries equation is
considered and the explicit expressions for the solutions are
obtained. Later in Section 3 conditions of chaos are derived on the
basis of studying subharmonic and homoclinic Melnikov functions. Then,
in Section 4, we transform our ODE (\ref{1.4}) into a mapping, for
which phase plots are presented together with the calculation of the
width of the stochastic layer. Section 5 contains Poincar\'e plots
numerically found in the vicinity of a saddle along with the Lyapunov
exponents which prove to be positive thereby supporting our conclusion
about chaotic behaviour in the system. Finally, Section 6 contains a summary and conclusions.
\section{Unperturbed case}
\setcounter{equation}{0}
\def\theequation{\arabic{section}.\arabic{equation}}
For the unperturbed system $(f_{0}=0)$ we simply have
\begin{equation}\label{2.1}
\beta u_{x x}=vu-\frac{u^2}{2}.
\end{equation}
This is an equation of motion for a nonlinear oscillator in a potential field
\begin{equation}\label{2.2}
P(u)=\frac{u^3}{6}-\frac{vu^{2}}{2}
\end{equation}
with the Hamiltonian
\begin{equation}\label{2.3}
H_{0}(u,z,x)=\frac{\beta z^2}{2}+\frac{u^3}{6}-\frac{vu^{2}}{2},
\end{equation}
If we now look for the solutions of (\ref{2.1}) in the form of elliptic Jacobi functions of modulus $m$:
\begin{equation}
u^{m}(x)=a+b\cn^{2}(kx\mid m),\mbox{ }0\leq m\leq 1,
\end{equation}
where $a$, $b$ and $k$ are depending on $m$ constants, then the following expressions can be easily obtained:
\begin{equation}\label{2.5}
\left\{
\begin{array}{l}
u^m(x)=v+\frac{v(1-2m)}{\sqrt{m^2-m+1}}+\frac{3vm}{\sqrt{m^2-m+1}} \cn^2(kx\mid m),
\\ z^m(x)=-\frac{6vmk}{\sqrt{m^2-m+1}} \cn(kx\mid m)\sn(kx\mid m)\dn(kx\mid m),
\\ T^m=\frac{2K(m)}{k},
\\ k\equiv \frac{1}{2}\sqrt{\frac{v}{\beta}} (m^2-m+1)^{-\frac{1}{4}},
\end{array}
\right.
\end{equation}
Here $T^m$ is the period of oscillations and $K(m)$ is the complete elliptic integral of the first kind. In the limit $m\to 1$ (\ref{2.5}) yields the separatrix
\begin{equation}\label{2.6}
\left\{
\begin{array}{l}
u_{0}(x)=3v\sech^2(\frac{1}{2}\sqrt{\frac{v}{\beta}}x),
\\ z_{0}(x)=-3v\sqrt{\frac{v}{\beta}}\sinh({\frac{1}{2}\sqrt{\frac{v}{\beta}}x}) \sech^3({\frac{1}{2}\sqrt{\frac{v}{\beta}}x}).
\end{array}
\right.
\end{equation}
Phase plot of the system (\ref{2.1}) as well as the potential energy (\ref{2.2}) is presented in Figure 1.

\section{Melnikov chaos}
\setcounter{equation}{0}
\def\theequation{\arabic{section}.\arabic{equation}}
The source of chaotic motion in our system is the homoclinic orbit on
a phase plane. In the unperturbed case this orbit is formed by
coinciding stable and unstable manifolds of the saddle.  When there is
a perturbation, the homoclinic orbit can be broken to yield the
transverse intersection of stable and unstable manifolds, which gives
rise to chaotic behaviour near the separatrix.
\begin{figure}
\hspace{4cm}
\epsfig{width=4cm,file=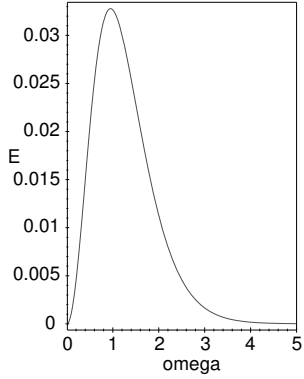}
\caption{The width of stochastic layer as the function of perturbation frequency (\ref{4.5}). Here $\beta=1$, $v=1$, $f_{0}=0.01$.}
\end{figure}
As far as one of the precursors of chaos is the appearance of subharmonics, we'll start with the subharmonic Melnikov function defined for the periodic orbits as
\begin{equation}\label{3.1}
M^{r/s}(\theta)=\int_{0}^{rT}{f(q^{m}(x))\wedge g(q^{m}(x+\theta))dx},
\end{equation}
where $f(q^{m})$ is the unperturbed vector field corresponding to the solution
$q^{m}(x)=(u^{m},z^{m})^T$ with the period $T^{m}=rT/s$ for $r$, $s$
relatively prime integers, $g$ is the small vector of perturbation,
and the wedge product is defined by $a\wedge b=a_{1}b_{2}-a_{2}b_{1}$.

In considering our problem we substitute the term $z^{m}(x)$ for $f$,
and the term of perturbation $f_{0}\cos\omega x$ for $g$ into the subharmonic Melnikov function, therefore obtaining:
\begin{equation}\label{3.2}
M^{r/s}(\theta)=\frac{f_{0}}{\beta}\int_{0}^{rT}{z^{m}(x)\cos\omega(x+\theta)dx}.
\end{equation}
Calculations with the help of Fourier representation of elliptic
functions \cite{AS} finally give:
\begin{equation}\label{3.3}
M^{r/s}(\theta)=\frac{3\pi rvmf_{0}}{\beta\sqrt{m^2-m+1}}b_{n}(m)\sin\omega\theta,
\end{equation}
where
\begin{equation}
b_{n}(m)=\frac{\pi^{2}(n+1/2)}{mK^{2}(m)\sinh\left[ \pi(n+1/2)\frac{K'(m)}{K(m)}\right]},\mbox{ }K'(m)=K(1-m).
\end{equation}
As it is clearly seen from (\ref{3.3}) $M^{r/s}(\theta)$ has simple
zeros, what proves the presence of subharmonic bifurcations \cite{GH}.

To move further the homoclinic Melnikov function $M(\theta)$ can be introduced as a limit of $M^{r/s}(\theta)$ when $s=1$, $r\rightarrow \infty$, and $m\rightarrow 1$:
\begin{equation}\label{3.5}
M(\theta)=\frac{f_{0}}{\beta}\int_{-\infty}^{\infty}{z_{0}(x)\cos\left[\omega (x+\theta)\right]dx}=\frac{12\pi\omega^{2}f_{0}}{\sinh{\pi\omega\sqrt{\frac{\beta}{v}}}}\sin{\omega\theta}.
\end{equation}
Surely this function also has simple zeros on $\theta$ and therefore
proves the existence of transverse homoclinic orbits. The presence of
these orbits means, via Smale-Birkhoff theorem, the appearance of a
Smale horseshoe in the vicinity of an unperturbed saddle and thereby
homoclinic chaos in this region of a phase space \cite{GH}. We note
here that unlike the situation with dissipation ( like in \cite{GT} or
\cite{CDPT}), when Melnikov theory resulted in some relation between
the perturbation amplitude, frequency and the dissipation coefficient,
in our case there are no such restrictions. It should be also noted
that although the presence of a stochastic layer near the separatrix
can be stated for rather generic hamiltonian
\begin{figure}
\hspace{1.5cm}
\epsfig{width=5cm,file=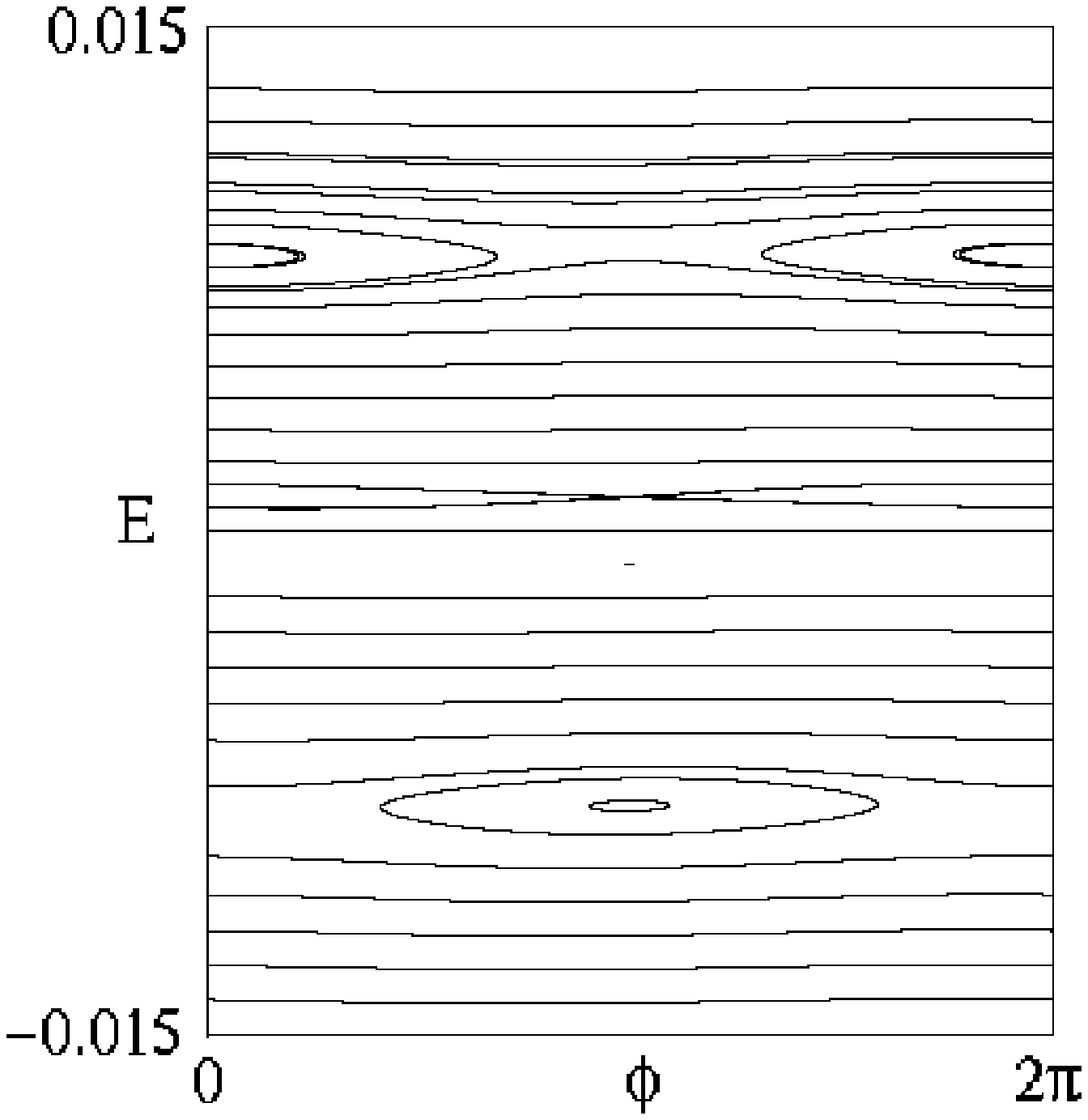}\hspace{2cm}
\epsfig{width=5cm,file=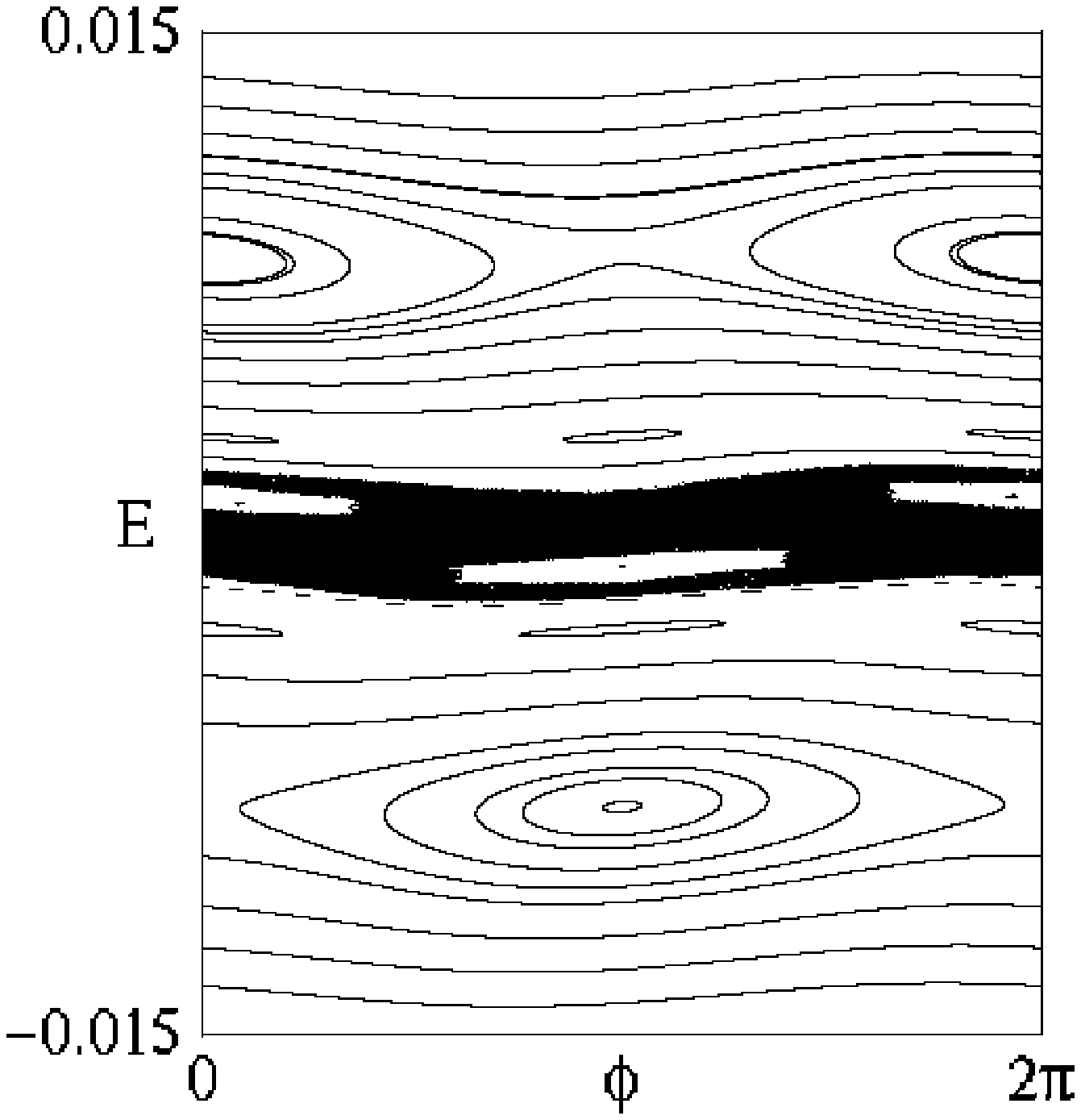}
\caption{Separatrix mapping (\ref{4.4}) with $v=1$, $\beta=1$, $\omega=3$. (Left) $f_{0}=1.83\cdot 10^{-3}$. (Right) $f_{0}=9.1\cdot 10^{-3}$.}
\end{figure}
\begin{figure}
\hspace{1.5cm}
\epsfig{width=5cm,file=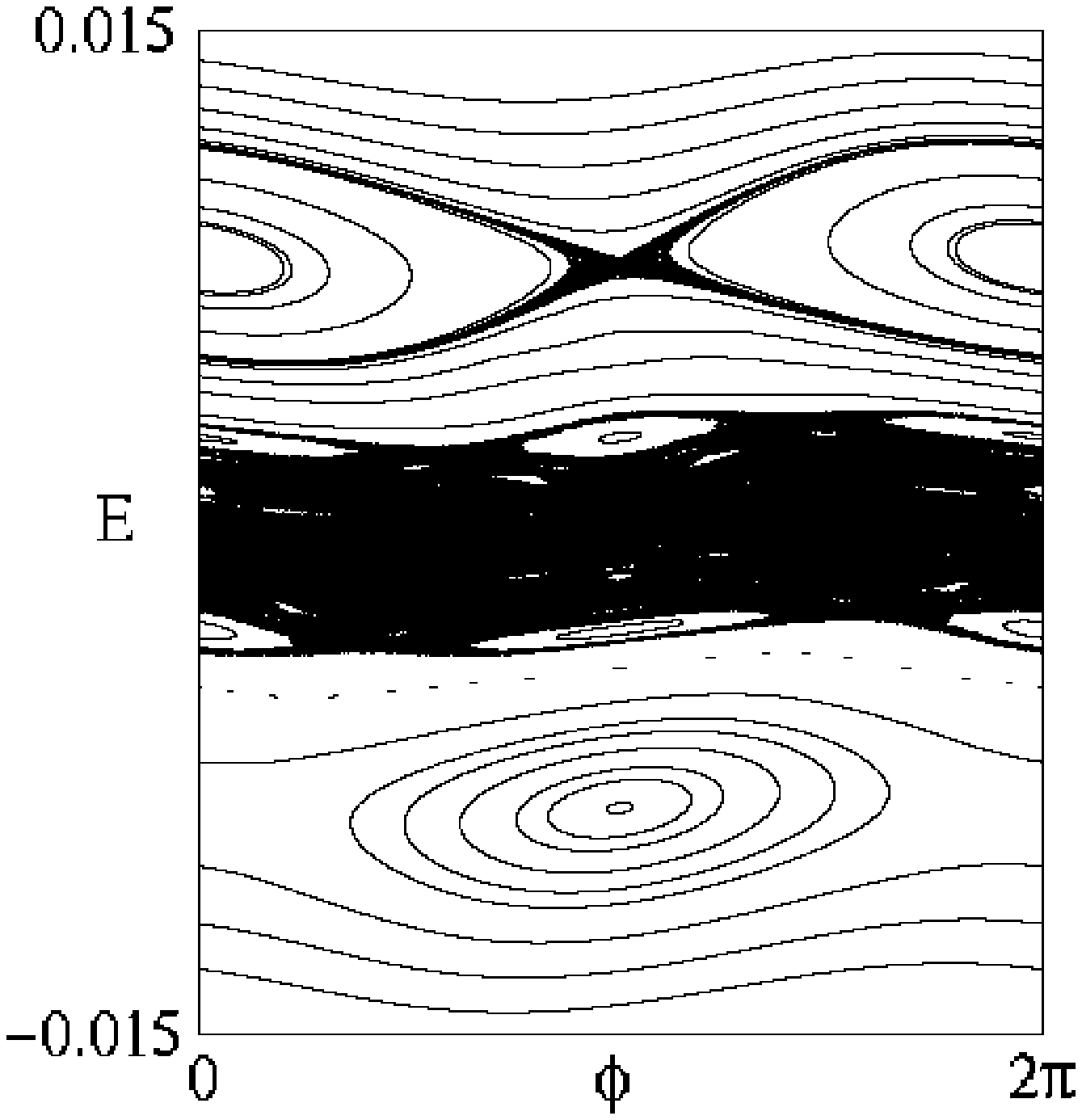}\hspace{2cm}
\epsfig{width=5cm,file=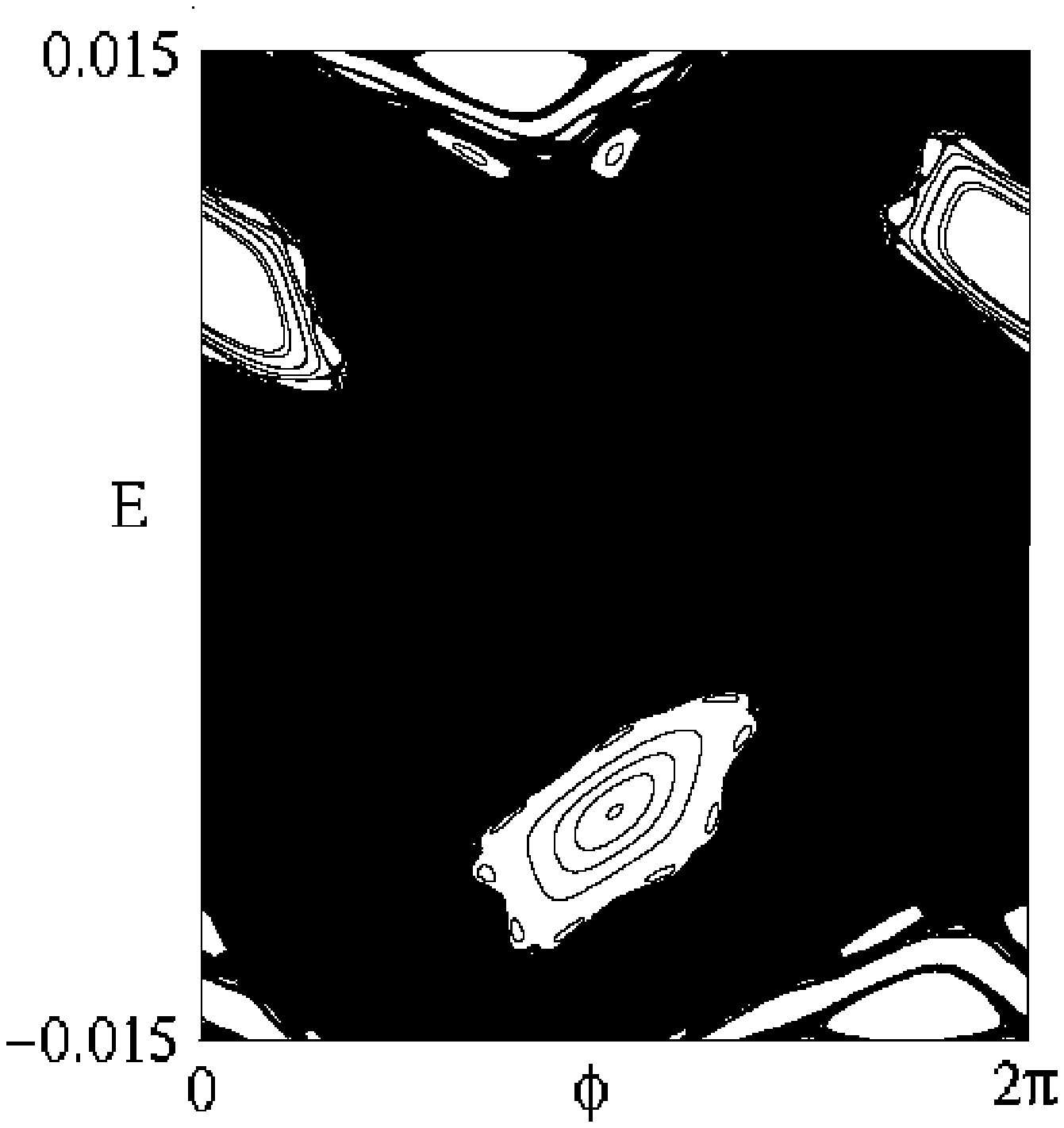}
\caption{Separatrix mapping (\ref{4.4}) with $v=1$, $\beta=1$, $\omega=3$. (Left) $f_{0}=1.83\cdot 10^{-2}$. (Right) $f_{0}=9.1\cdot 10^{-2}$.}
\end{figure}
perturbation making system near-integrable \cite{Za}, the transverse intersection of stable and unstable manifolds is still a question which should solved by Melnikov theory in every particular case.

\section{Near-separatrix motion}
\setcounter{equation}{0}
\def\theequation{\arabic{section}.\arabic{equation}}
To study the chaotic motion near the separatrix we shall use the whisker mapping (separatrix mapping) \cite{W}. To derive it we note first that the change in the unperturbed energy is given by:
\begin{equation}
\frac{dH_{0}}{dt}=-\frac{\partial H_{0}}{\partial p}\frac{\partial V}{\partial x}=\beta f_{0}z(x)\cos{\omega x},
\end{equation}
where $p=\beta z$ is the momentum. Since the perturbation is a periodic function of time, we may introduce the phase of perturbation:
\begin{equation}
\phi=\omega x+\const,
\end{equation}
which is an equivalent of time. In order to obtain the desired
separatrix mapping one has to consider the discretized time scale
$x_{n}$, and the mapping will involve $\phi_{n}$ and $E_{n}$ as
variables. The change of phase after one period of oscillations is
equal to $\omega T^{m}(E_{n+1})$, while energy change $\Delta E_{n+1}=E_{n}+\Delta E_{n}$. Now we can estimate the change in the unperturbed energy $\Delta E_{n}$ per period of motion in proximity separatrix at time $x_{n}$:
\begin{equation}\label{4.2}
\Delta E_{n}=\beta f_{0}\int_{x_{n}-T/2}^{x_{n}+T/2}{z(x)\cos\left[\omega(x+x_{n})\right]dx}\approx\beta f_{0}\int_{-\infty}^{\infty}{z_{0}(x)\cos\left[\omega(x+x_{n})\right]dx}.
\end{equation}
Here we have approximated $\Delta E_{n}$ by evaluating the integral (\ref{4.2}) on the unperturbed separatrix in accordance with the standard procedure \cite{Za}. As it is clearly seen from the comparison of (\ref{3.5}) and (\ref{4.2}), $\Delta E_{n}=\beta^{2} M(x_{n})$. Expansions of $T^{m}$ and $E(m)$ in the vicinity of $m=1$ give finally the following correlation between them: $T(E)\approx 2\sqrt{(v/\beta)}\ln\left[24/\sqrt{2\mid E\mid}\right]$. Using this approximate period we can write the separatrix mapping in the following form:
\begin{equation}\label{4.4}
\left\{
\begin{array}{l}
E_{n+1}=E_{n}+\frac{12\pi f_{0}\omega^{2}\beta^{2}}{\sinh\pi\omega\sqrt{\frac{v}{\beta}}}\sin\phi_{n},\\
\phi_{n+1}=\phi_{n}+2\sqrt{\frac{v}{\beta}}\omega\ln\frac{24}{\sqrt{2\mid E_{n+1}\mid }} \mbox{ ( mod }2\pi\mbox{ )}.
\end{array}
\right.
\end{equation}
If to measure the stretching of a small phase interval with the parameter $K=|\delta\phi_{n+1}/\delta\phi_{n}-1|$ \cite{Za}, one can detect the border of chaotic motion as an appearance of a local instability in phase: $K\geq1$. For the mapping (\ref{4.4}) this condition gives the following approximation for the width of the stochastic layer $E_{st}$:
\begin{equation}\label{4.5}
\mid E\mid\leq E_{st}=\sqrt{\frac{v}{\beta}}\frac{12\pi f_{0}\omega^{3}\beta^{2}}{\sinh\pi\omega\sqrt{\frac{v}{\beta}}}.
\end{equation}
Typical form of this dependence is presented in the Fig. 2. In Fig. 3,
4 we have plotted the phase plane of the mapping ($E$, $\phi$) for different
values of the perturbation amplitude $f_{0}$. Fig. 3(Right) shows how the
stochastic layer arises around the unperturbed separatrix. Then the
width of this layer increases together with $f_{0}$, leading finally
to a chaotic sea, as demonstrated in Fig. 4(Right).

\section{Numerical simulations}
\setcounter{equation}{0}
\def\theequation{\arabic{section}.\arabic{equation}}
\begin{figure}
\hspace{2cm}
\epsfig{width=4cm,file=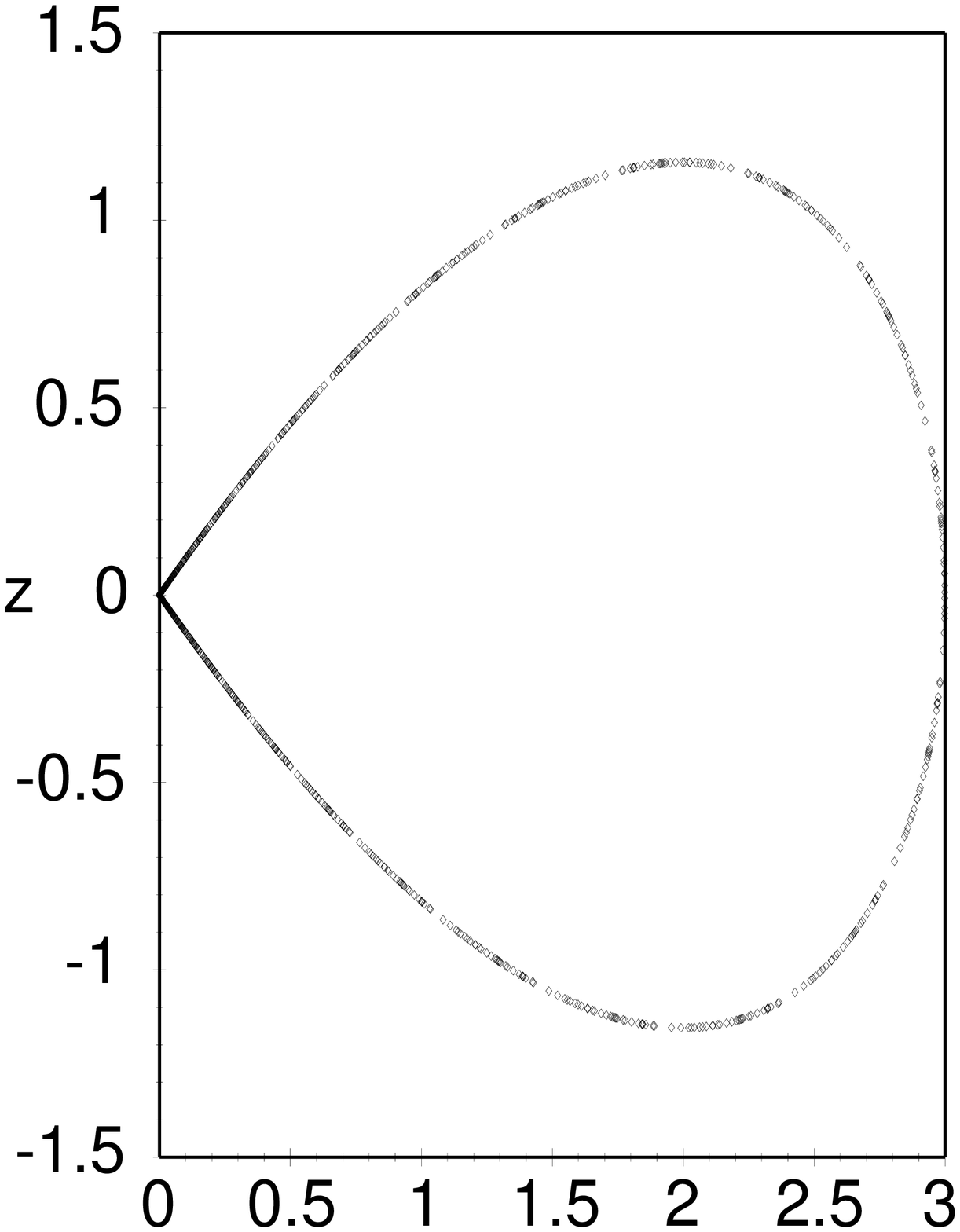}\hspace{2cm}
\epsfig{width=4cm,file=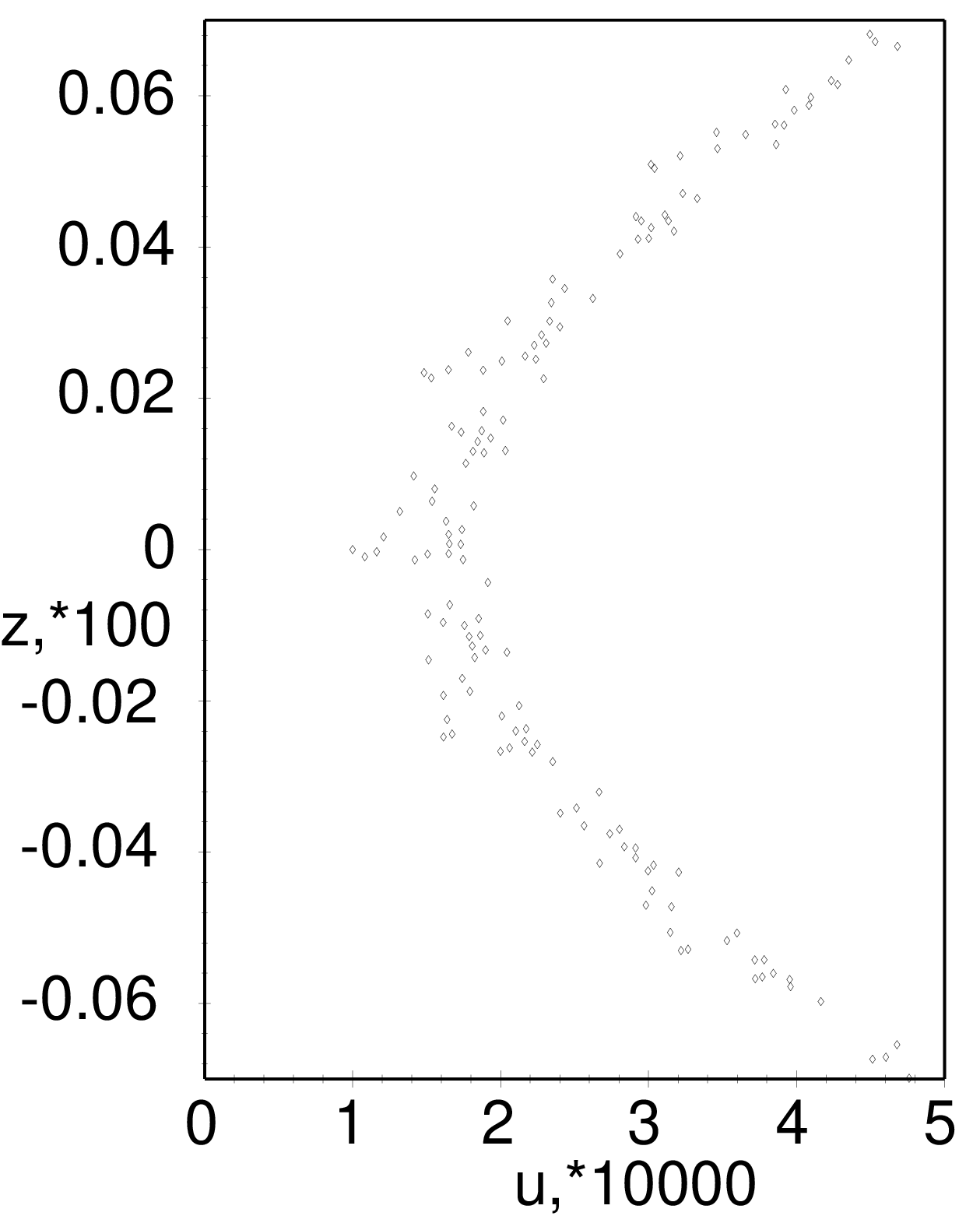}
\caption{(Left) Poincar\'e section of the system (\ref{1.4}). (Right) The same near the saddle (0,0). Here $f_{0}=0.1$, $\omega=15$, $v=1$, $\beta =1$, and initial conditions are taken in the point (0.0001,0).}
\end{figure}
\begin{figure}
\hspace{2cm}
\epsfig{width=4cm,file=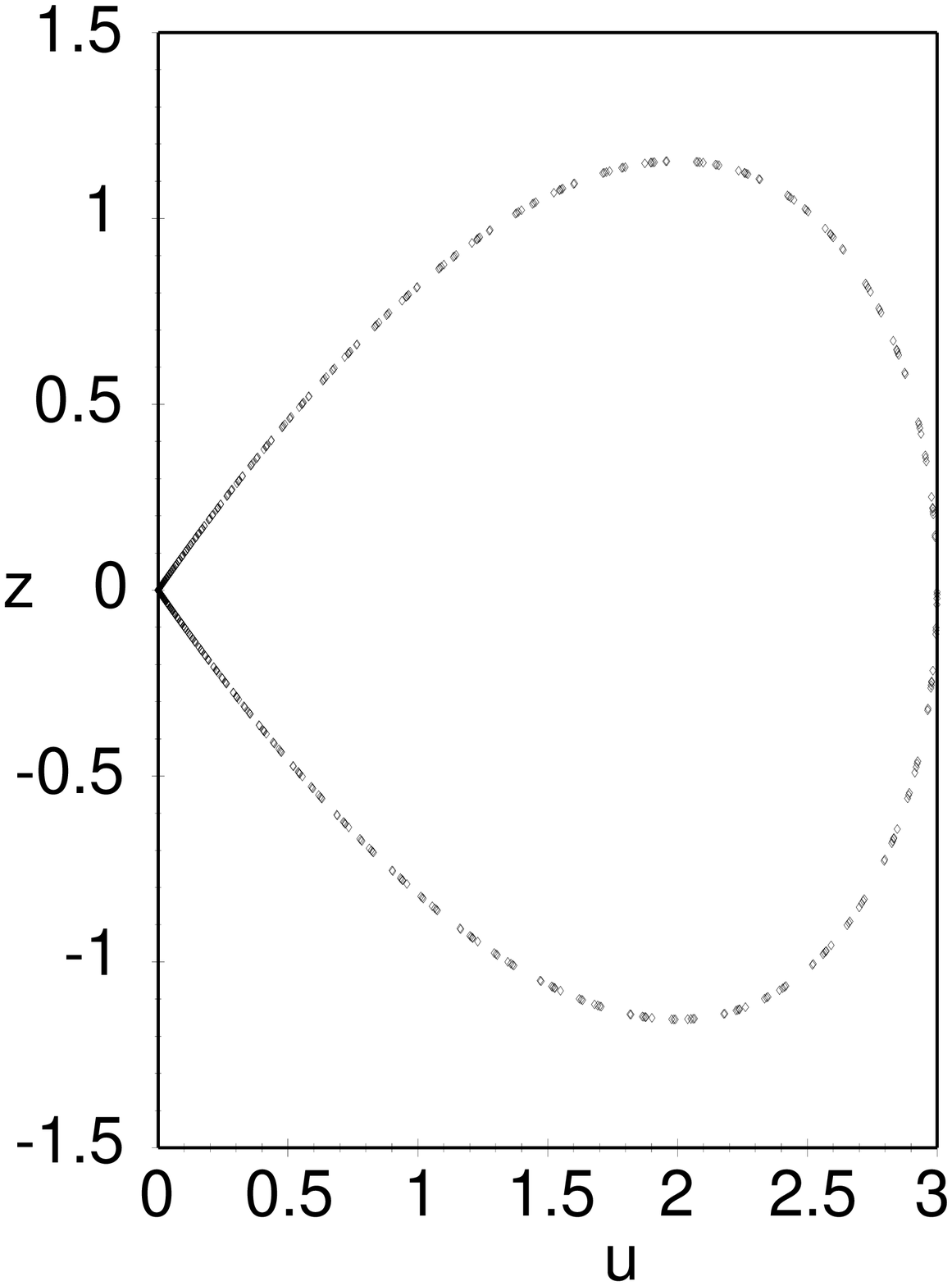}\hspace{2cm}
\epsfig{width=4cm,file=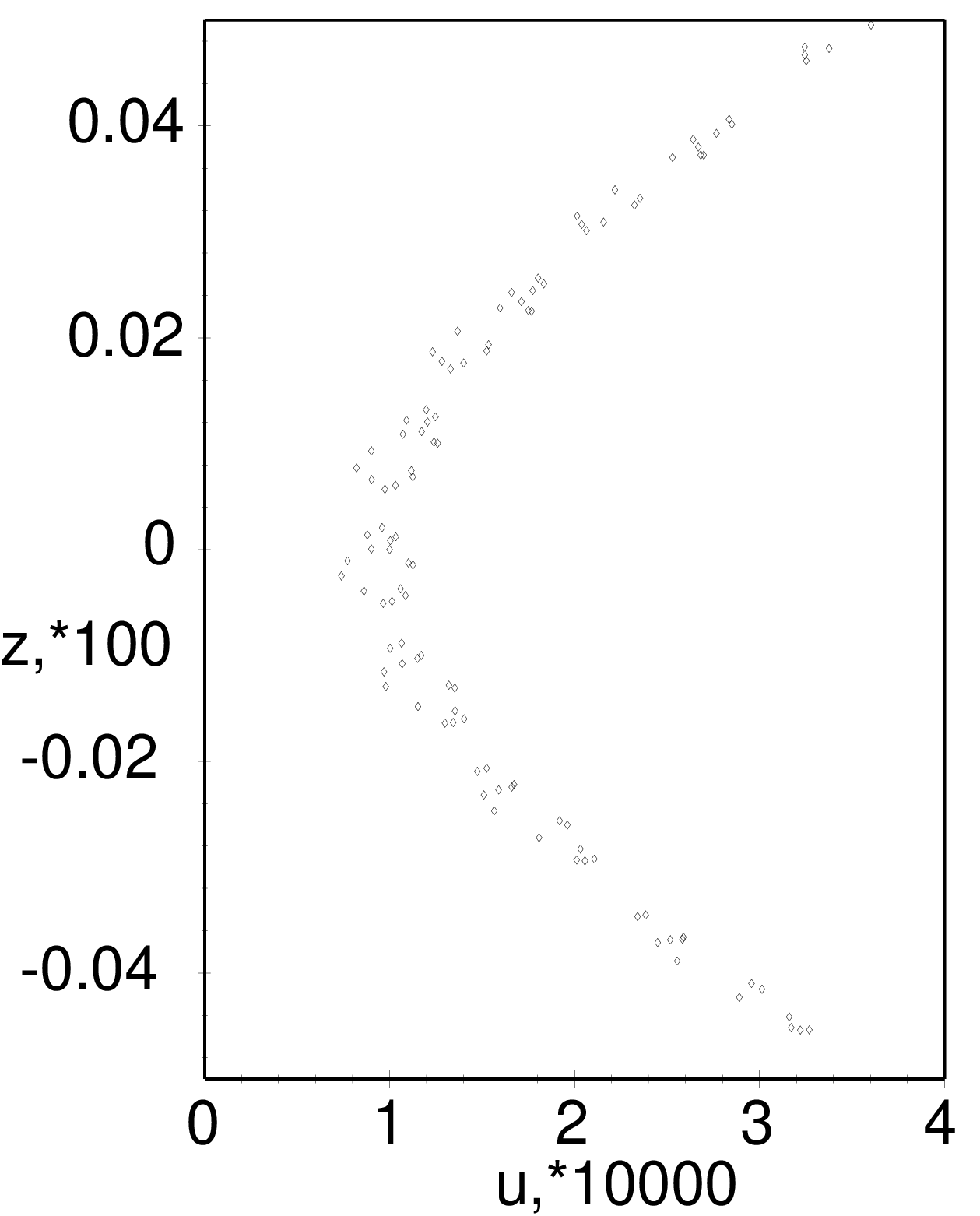}
\caption{ The same as in Figure 5 with $\omega=20$}
\end{figure}
In this section we present the Poincar\'e sections of (\ref{1.4}) in the vicinity of the saddle $(0,0)$, which show homoclinic crossings in this region. The methods used were Runge-Kutta-Fehlberg and Adams-Bashford-Moulton ("predictor-corrector") method.

Local chaotic behaviour is detected by the numerical calculation of the dominant Lyapunov exponent. It was computed using the method due to \cite{HR}. The system of equations, which defines the value of this exponent, was integrated with the Runge-Kutta fourth-order algorithm, while the equation (\ref{1.4}) for a fiducial trajectory was run with fourth-order symplectic scheme \cite{FR}.
In Figures 5-8 the Poincar\'e sections are presented for the system (\ref{1.4}) with the forcing amplitude $f_{0}=0.1$ and different values of the frequency: $\omega=15$, $20$, $25$ and $30$. The value of the largest Lyapunov exponent in these cases is $\lambda=0.107$, $\lambda=0.149$, $\lambda=0.169$ and $\lambda=0.188$ respectively, and so we can state that we really have chaotic motion in a close vicinity of a separatrix, as it was predicted. Numerical calculation of the Lyapunov exponent in the region of regular dynamics with initial conditions at $(1.5,0)$ yields $0.0001$, and therefore our results can be taken as significant.

\section{Conclusions}
In this work we have considered the stationary wave reduction of the
Korteweg-de Vries equation under small hamiltonian perturbations.
\begin{figure}
\hspace{2cm}
\epsfig{width=4cm,file=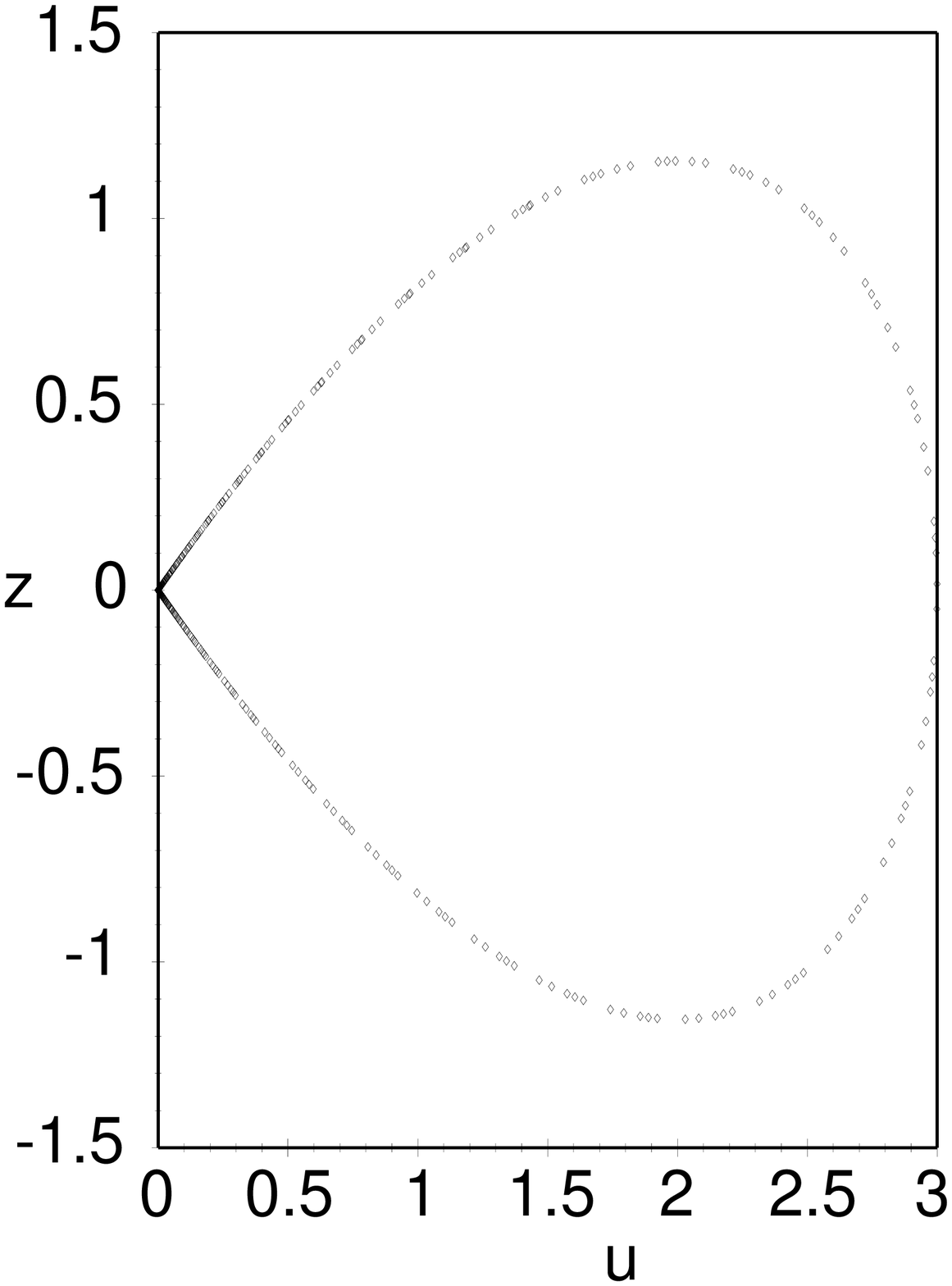}\hspace{2cm}
\epsfig{width=4cm,file=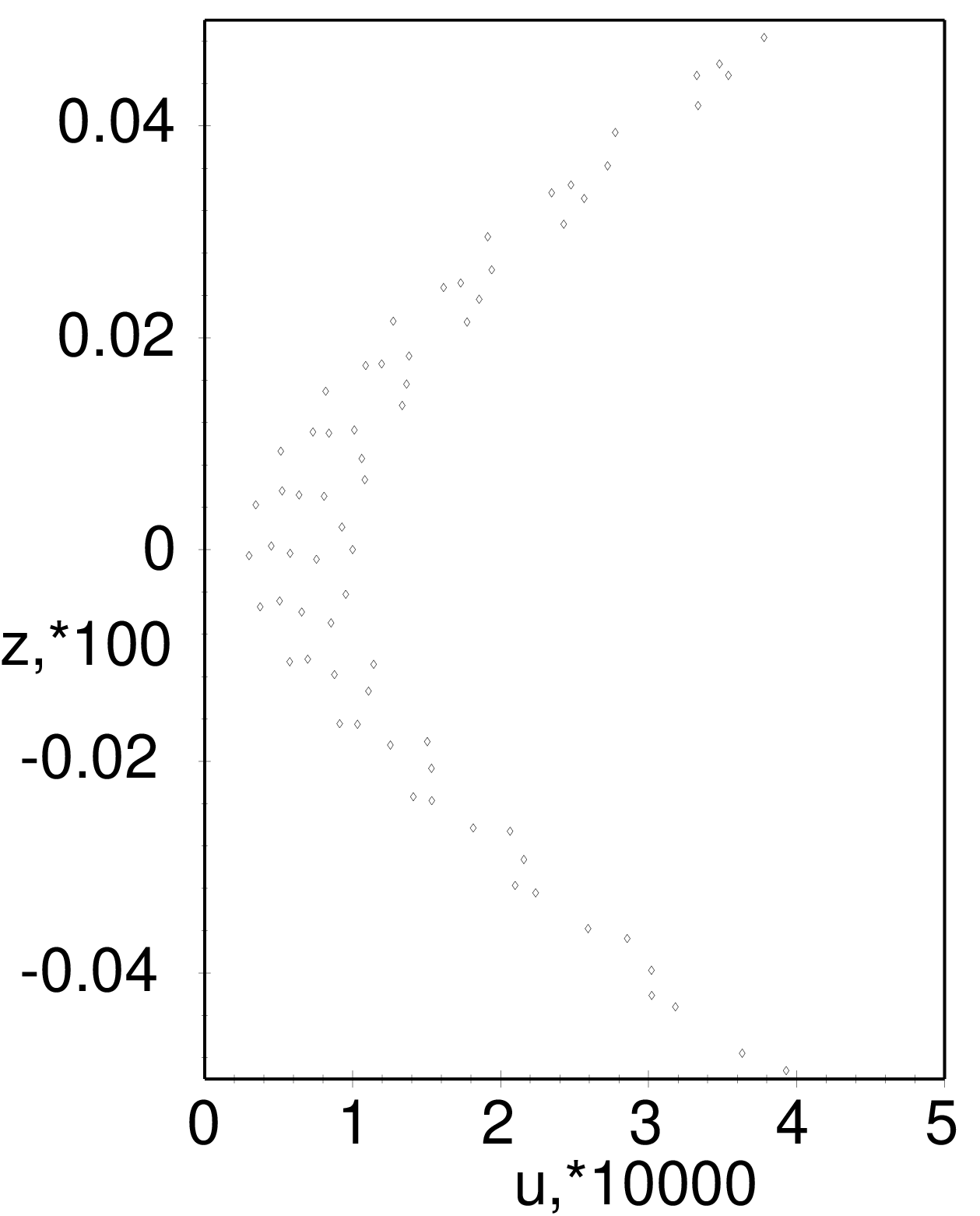}
\caption{ The same as in Figure 5 with $\omega=25$}
\end{figure}
\begin{figure}
\hspace{2cm}
\epsfig{width=4cm,file=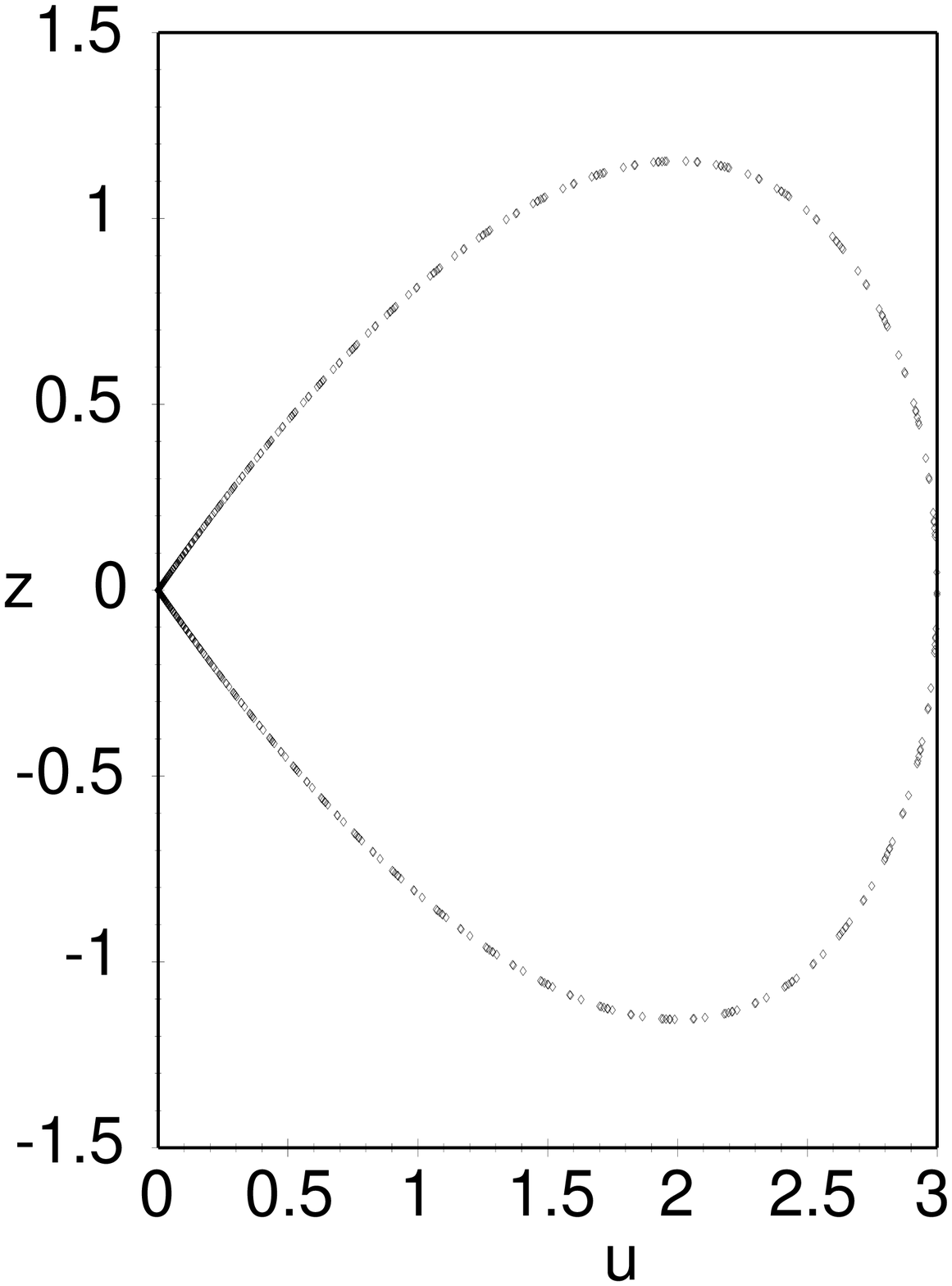}\hspace{2cm}
\epsfig{width=4cm,file=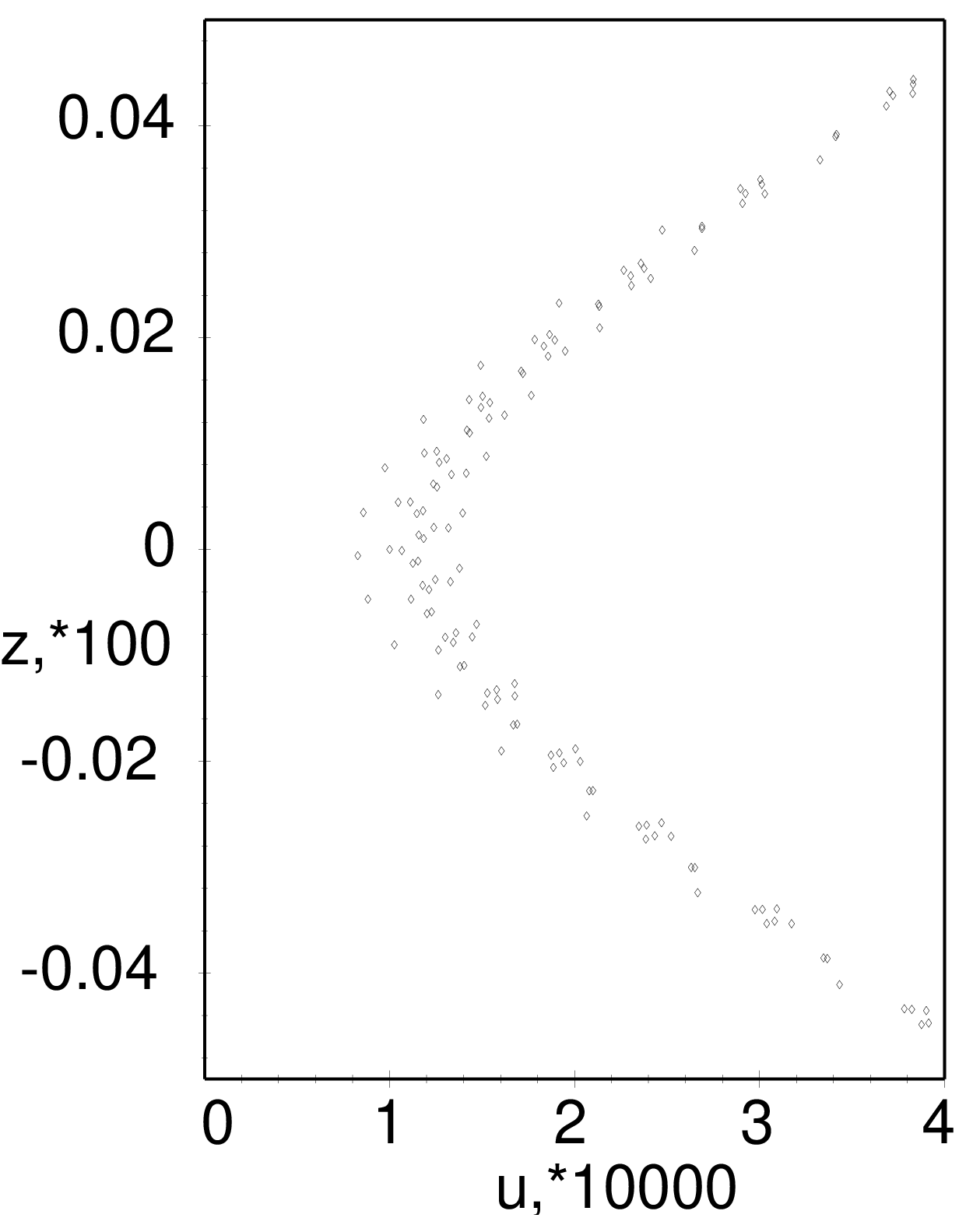}
\caption{ The same as in Figure 5 with $\omega=30$}
\end{figure}
On the basis of Melnikov theory it was shown how chaos can occur in this
situation via the appearance of subharmonics and a further homoclinic
tangle. Theoretical predictions about chaotic behaviour are supported
numerically by plotting Poincar\'e sections and calculation of the
corresponding Lyapunov exponents. In a near-separatrix region the
governing ODE was transformed into a mapping what allowed one to find
the width of the stochastic layer.

Our future work will focus on the development of kinetic description
of chaotic behaviour for the obtained mapping \cite{Za}, and with the
derivation of new methods for the symplectic integration of the KdV equation
under hamiltonian perturbations following the framework of \cite{LR},
where the same was done for sin-Gordon system.

\section*{Acknowledgments}
Author would like to thank Prof. M. Bestehorn for stimulating
discussions, Kai Neuffer for help with numerics, and Dr. S. Reich for
the information about symplectic integration methods.

\end{document}